\documentclass[aps,prl,superscriptaddress,amsmath,amssymb,reprint,showkeys,footinbib,preprintnumbers]{revtex4-2}

\usepackage{graphicx}
\usepackage{dcolumn}
\usepackage[normalem]{ulem}
\usepackage{float}
\usepackage{color}
\usepackage{xcolor}
\usepackage{bbm}

\usepackage[utf8]{inputenc}
\usepackage[T1]{fontenc}

\usepackage{hyperref}
\hypersetup{colorlinks=true,
linkcolor=[rgb]{0.94,0.19,0.22},
urlcolor=[rgb]{0.77,0.29,0.55},
citecolor=[rgb]{0,0.6,0.6}}

\begin{document}

\preprint{TIFR/TH/24-4}

\title{Large deviations of current for the symmetric simple exclusion process on a semi-infinite line, and on an infinite line with a slow bond}

\author{Kapil Sharma}
\email{kapil.sharma@tifr.res.in}
\affiliation{Department of Theoretical Physics, Tata Institute of Fundamental Research, Homi Bhabha Road, Mumbai 400005, India}

\author{Soumyabrata Saha}
\email{soumyabrata.saha@tifr.res.in}
\affiliation{Department of Theoretical Physics, Tata Institute of Fundamental Research, Homi Bhabha Road, Mumbai 400005, India}

\author{Sandeep Jangid}
\email{sandeep.jangid@tifr.res.in}
\affiliation{Department of Theoretical Physics, Tata Institute of Fundamental Research, Homi Bhabha Road, Mumbai 400005, India}

\author{Tridib Sadhu}
\email{tridib@theory.tifr.res.in}
\affiliation{Department of Theoretical Physics, Tata Institute of Fundamental Research, Homi Bhabha Road, Mumbai 400005, India}

\date{\today}

\begin{abstract}
Two influential exact results in classical one-dimensional diffusive transport are about current statistics for the symmetric simple exclusion process: one in the stationary state on a finite line coupled with two unequal reservoirs at the boundaries, and the other in the non-stationary state on an infinite line. We present the corresponding result for the intermediate geometry of a semi-infinite line coupled with a single reservoir. This result is obtained using the fluctuating hydrodynamics approach of macroscopic fluctuation theory and confirmed by rare event simulations using a cloning algorithm. We apply our exact result for solving several related challenging problems, namely, the full counting statistics in presence of a defect bond, exclusion process with localized injection, survival of a tagged particle in presence of an absorbing boundary, and the stretched exponential decay in a kinetically constrained model. 
\end{abstract}


\maketitle

Current fluctuations in non-equilibrium transport have long been a subject of interest in both classical and quantum contexts~\cite{2002_Prähofer_Current,2005_Bertini_Current,*2006_Bertini_Non,2007_Derrida_Non,2015_Lazarescu_The,2020_Banerjee_Current,2023_Bello_Current,1995_Lee_Universal,2023_McCulloch_Full,2024_Wienand_Emergence,turkeshi2024}. The major interest in these investigations lies in the full counting statistics in terms of large deviations of current. Apart from being a contender for extension of free energy for non-equilibrium systems, large deviation function (ldf) in general can characterize various peculiarities of non-equilibrium conditions, such as non-local response, emergent symmetries, and low-dimensional phase transitions~\cite{2005_Bodineau_Distribution,2007_Derrida_Non,2008_Bodineau_Long,2010_Jona_From,2009_Touchette_The,2009_Bertini_Towards,2011_Hurtado_Symmetries,2018_Baek_Dynamical}. However, estimating ldf poses a major challenge, often requiring specialized integrability techniques~\cite{2001_Schutz_Exactly,2015_Lazarescu_The,2015_Mallick_The} tailored to curated models or clever numerical sampling schemes of rare events~\cite{2006_Giardinà_Direct,2007_Lecomte_Numerical,2011_Giardinà_Simulating,2019_Perez_Sampling,Hartmann_computer_sim}. Understandably, exact results about ldf play an important role in the landscape of non-equilibrium physics, providing a benchmark for affirming qualitative predictions of approximate methods. Our work in this \emph{Letter} presents a non-trivial addition to this list of exact results, fostering solution to further problems of practical significance~\cite{1989_Spohn_Stretched,2012_Krapivsky_Symmetric,2013_Franco_Phase,*2016_Franco_Phase}.

Among the widely studied stochastic models of classical transport, is the symmetric simple exclusion process (SSEP)~\cite{1991_Spohn_Large,1999_Kipnis_Scaling,1999_Liggett_Stochastic,2007_Derrida_Non}. SSEP, along with its driven variants, has attained the status of a paradigmatic model in non-equilibrium statistical physics~\cite{2011_Chou_Non,2015_Mallick_The}. Two celebrated exact results for SSEP concern the large deviations of current on a finite lattice coupled with two unequal reservoirs, and on an infinite lattice starting with a non-stationary state. The two geometries represent distinct non-equilibrium scenarios: a stationary state for finite systems, and a time-dependent state relaxing towards an asymptotic equilibrium state for infinite systems. They illuminate crucial differences in their fluctuations. The finite system, in the long run, holds no memory of the initial state, while the infinite system exhibits an unusual dependence on the initial state even at long times~\cite{2009_Derrida_Current2,2012_Krapivsky_Fluctuations}.

The large deviations of current in these two geometries were obtained using the additivity conjecture~\cite{2004_Bodineau_Current,2007_Bodineau_Cumulants,2021_Derrida_Large} and integrability methods, such as the diagonalization of tilted matrix~\cite{2004_Derrida_Current} or the matrix product states~\cite{2012_Gorissen_Exact,2015_Lazarescu_The} for the finite lattice, and the Bethe ansatz~\cite{2009_Derrida_Current} for the infinite lattice. These microscopic results were subsequently verified~\cite{2005_Bertini_Current,*2006_Bertini_Non,2022_Mallick_Exact,*2024_Mallick_Exact,2024_Saha_Large} using a fluctuating hydrodynamics framework~\cite{2009_Derrida_Current,2015_Bertini_Macroscopic,2023_Jona_Current}. This framework, now famously known as the macroscopic fluctuation theory (MFT)~\cite{2001_Bertini_Fluctuations,*2002_Bertini_Macroscopic,2005_Bertini_Current,*2006_Bertini_Non}, emerged in the early 2000s from the seminal works of Bertini, De Sole, Gabrielli, Jona-Lasinio, and Landim, which presented a general approach for characterizing non-equilibrium fluctuations of diffusive systems. MFT has successfully led to exact results of large deviations in exclusion processes~\cite{2005_Bertini_Current,*2006_Bertini_Non,2007_Tailleur_Mapping,*2008_Tailleur_Mapping,2015_Bertini_Macroscopic,2022_Mallick_Exact,2024_Saha_Large} and related transport models~\cite{2015_Bertini_Macroscopic,2022_Bettelheim_Inverse,*2022_Bettelheim_Full,2023_Agranov_Macroscopic,2024_Bettelheim_Complete}.

In this \emph{Letter}, we consider the intermediate scenario: SSEP on a semi-infinite lattice~\cite{1975_Liggett_Ergodic,2004_Grosskinsky_Phase} coupled to a boundary reservoir with a density that is different from the initial bulk density of the system. This elucidates a non-equilibrium regime evolving into an asymptotic state in equilibrium with the reservoir~\cite{2023_Saha_Current}. In fact, the current statistics in this geometry describes the early time ($t\ll L^2$) statistics near the reservoir of a finite system of length $L$, extending the previous steady state ($t\gg L^2$) result \cite{2004_Bodineau_Current}. The statistics have also been useful recently in the context of Mpemba effect~\cite{turkeshi2024} in quantum circuits and particle injection in a lattice~\cite{2014_Krapivsky_Lattice, 2012_Krapivsky_Symmetric}. We shall further see that the semi-infinite geometry has direct relation to transport in presence of defect and survival statistics of tracer. Despite the relevance, only limited results~\cite{2012_Williams_Combinatorics,2013_Tracy_The,2018_Duhart_The,2023_Saha_Current} are known for this intermediate geometry, particularly because extending the aforementioned integrability methods for this geometry proves challenging. Even a solution via MFT has remained elusive~\cite{2023_Saha_Current}.

We overcame these technical challenges by uncovering a novel mapping to the infinite-line problem that leads to an exact result for the ldf of current for SSEP on the semi-infinite line, which is our main result. The mapping to the infinite-line-problem is possible due to certain symmetries of the Euler-Lagrange equations and the boundary conditions of the associated variational problem within MFT. These techniques, however, could not be obviously exploited within microscopic approaches. Our exact result is a testament to the power of MFT for non-equilibrium systems that are otherwise formidable to approach.

Our exact result for the semi-infinite SSEP enables the solution of three additional challenging problems. First, we obtain the exact ldf of current for an infinite SSEP with a localized slow bond, extending the celebrated result of Derrida and Gerschenfeld \cite{2009_Derrida_Current}. Second, we derive the corresponding ldf for fast particle injection at a single site~\cite{2014_Krapivsky_Lattice, 2012_Krapivsky_Symmetric} in an infinite SSEP. Third, we compute the survival probability of a tracer in a fully packed SSEP with an absorbing boundary. This result extends the classical static-target~\cite{2014_Meerson_Survival} survival problem, which concerns the special case where no particle escapes the system.

\begin{figure}
\centering
\includegraphics[width=0.95\linewidth]{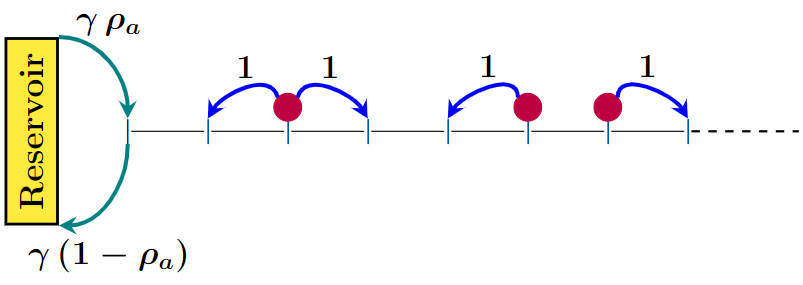}
\caption{SSEP on a semi-infinite lattice coupled with a reservoir of density $\rho_a$ with a coupling strength $\gamma>0$.}
\label{fig: dynamics}
\end{figure}

\textit{Model and main result}: SSEP on a semi-infinite lattice (see Fig.~\ref{fig: dynamics}), with sites indexed by $i=1,2,\cdots$, is composed of continuous-time hard-core random walkers hopping to adjacent sites with unit rate provided the target site is empty. At the boundary site $i=1$, particles are injected following exclusion with rate $\gamma\rho_a$ and removed at rate $\gamma(1-\rho_a)$, which models~\cite{2007_Derrida_Non} coupling with a reservoir of density $\rho_a$. For each site $i$ at a given time $\tau$, we assign a binary occupation number $n_i(\tau)$ that takes values $0$ and $1$, depending on whether the site is empty or occupied, respectively. Initially, sites are filled following Bernoulli distribution with a uniform average density $\rho_b$. 
Our main result concerns the time-integrated current, $Q_T$, which represents the total flux of particles from the reservoir into the system over a time-period $T$. In the hydrodynamic description~\cite{SM}, expressed in terms of coarse-grained density $\rho(i/\sqrt{T},\tau/T)\simeq n_i(\tau)$, the flux 
\begin{equation}\label{eq:QT}
Q_T={\sqrt{T}}\int_0^\infty\mathrm{d}x\,[\rho(x,1)-\rho(x,0)]
\end{equation}
measures the net change in the number of particles in the system.
In the large $T$ limit, its generating function has the asymptotics
\begin{equation}
\big<\mathrm{e}^{\lambda Q_T}\big>\sim\mathrm{e}^{\sqrt{T}\mu_{\text{si}}(\lambda,\rho_a,\rho_b)}. \label{mom_gen_fun}
\end{equation}
The scaled cumulant generating function (scgf) \begin{subequations}\label{musi_complete} 
\begin{equation}\label{mu_si}
\mu_{\text{si}}(\lambda,\rho_a,\rho_b)=R_{\text{si}}\big(\omega(\lambda,\rho_a,\rho_b)\big)
\end{equation}
(subscript `$\text{si}$' denotes semi-infinite) with
\begin{equation}\label{PW_Rsi}
R_{\text{si}}(\omega)=
\begin{cases}
R(\omega) & \text{for }\omega\ge-\frac{1}{2}\\
\displaystyle{-R(\omega)-\frac{1}{\sqrt{\pi}}\zeta\bigg(\frac{3}{2}\bigg)} & \text{for }\omega\le-\frac{1}{2}
\end{cases}
\end{equation}
where $\zeta(z)$ is the Riemann zeta function and
\begin{equation}\label{IntegralRep}
R\big(\omega\big)=\int_{-\infty}^\infty\frac{\mathrm{d}k}{2\pi}\log{\big[1+4\,\omega\,(1+\omega)\,\mathrm{e}^{-k^2}\big]}
\end{equation}
and $\omega$ is a function of the parameters defined as~\cite{2009_Derrida_Current,2012_Krapivsky_Fluctuations}
\begin{equation}\label{ssep_parameter}
\omega(\lambda,\rho_a,\rho_b)=\rho_a(1-\rho_b)(\mathrm{e}^{\lambda}-1)+\rho_b(1-\rho_a)(\mathrm{e}^{-\lambda}-1).
\end{equation}
\end{subequations}
This parametric dependence of scgf on $\lambda$, $\rho_a$, and $\rho_b$ through a single function $\omega$ arises from a symmetry of the underlying dynamics of SSEP~\cite{2004_Derrida_Current,2009_Derrida_Current,2010_Lecomte_Current}. The piecewise function \eqref{PW_Rsi} is constructed by analytic continuation of $R(\omega)$ demanding a convex scgf independently confirmed by numerical simulation (see End Matter). The expression~\eqref{musi_complete} is consistent with an earlier result~\cite{2023_Saha_Current} in the low density limit. 

The function $R(\omega)$ has different representations. For example, $R(\omega)=-1/(2\sqrt{\pi})\mathrm{Li}_{3/2}(-4\omega(1+\omega))$ in terms of the poly-logarithm function $\mathrm{Li}_s(z)$ is similar to the expression of the number of particles in Fermi-Dirac statistics. Another representation in terms of Fredholm determinant
\begin{equation}
R(\omega)=\frac{1}{2\pi}\log{\det{[\mathbbm{1}+4\omega(1+\omega)K(x,y)]}},
\end{equation}
with the kernel $K(x,y)=\delta(x-y)\exp{(\partial_x^2)}$, finds analogue in the current statistics for asymmetric exclusion process~\cite{2000_Johansson_Shape,2008_Tracy_A,2009_Tracy_Total}.

It is instructive to compare~\eqref{musi_complete} with the corresponding result for the infinite line~\cite{2009_Derrida_Current}. Similar to the infinite line case~\cite{2009_Derrida_Current2}, the scgf~\eqref{musi_complete} admits the Gallavotti-Cohen-type fluctuation symmetry \cite{Evans1993PRL,Gallavotti1995PRL,Kurchan1998JPA,Lebowitz1999JSP}
\begin{equation}\label{gallavotti_cohen}
\mu_{\text{si}}(\lambda,\rho_a,\rho_b)=\mu_{\text{si}}\bigg(\log{\frac{\rho_b\,(1-\rho_a)}{\rho_a\,(1-\rho_b)}}-\lambda,\rho_a,\rho_b\bigg).
\end{equation}

\begin{figure}
\centering
\includegraphics[width=\linewidth]{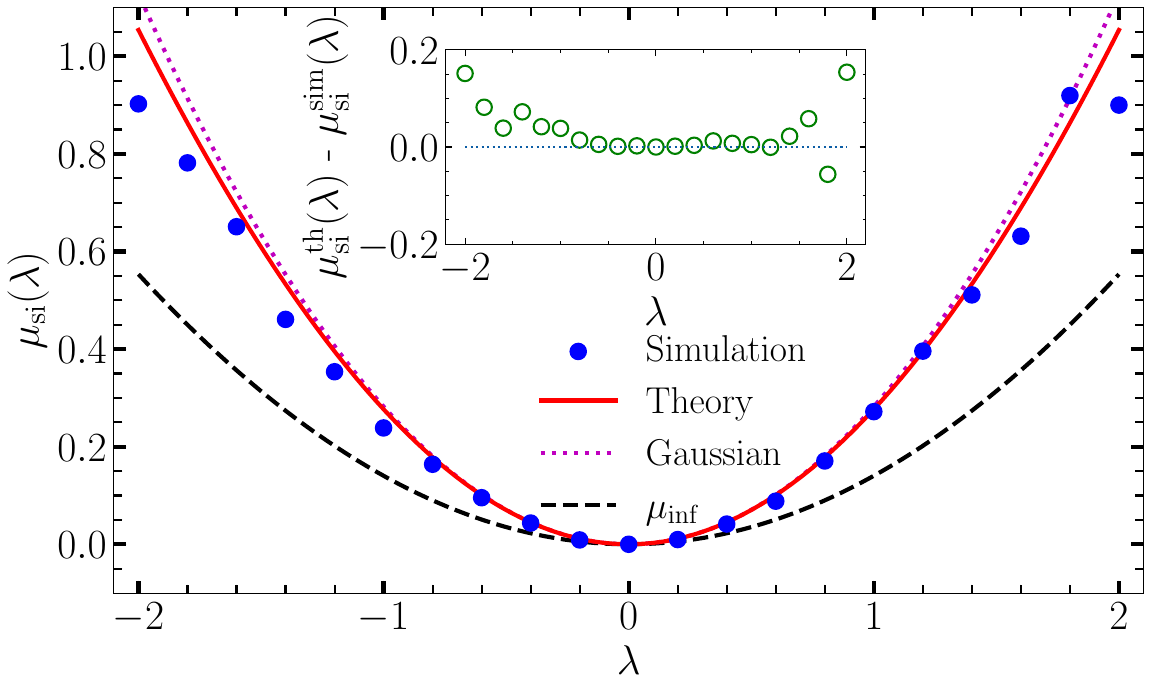} \caption{\textit{Scgf for semi-infinite SSEP:} Solid red line represents the theoretical result of scgf~\eqref{musi_complete} for $\rho_{a(b)}=0.5$, with blue dots representing the corresponding simulation result obtained by the cloning algorithm. The results closely match in the range $-1\leq\lambda\leq1$, with the deviations shown in the inset. For comparison, Gaussian approximation and the scgf for infinite line $\mu_{\text{inf}}$~\cite{2009_Derrida_Current} are shown in dotted magenta and black dashed line, respectively.}
\label{fig: Cloning Algorithm}
\end{figure}
\textit{Numerical confirmation}:
For confirming our result~\eqref{musi_complete} based on the hydrodynamic theory for SSEP, we have independently generated the scgf using rare-event simulation based on a continuous-time cloning algorithm~\cite{SM,2006_Giardinà_Direct,2007_Lecomte_Numerical} with $10^5$ clones and measured over duration $T=250$. The data, plotted in Fig.~\ref{fig: Cloning Algorithm}, shows a good agreement with our theoretical result~\eqref{musi_complete} for a reasonably large parameter range of $\lambda$. Deviations emerging at larger values of $\lambda$ are a consequence of finite-size effects~\cite{SM}.

\textit{LDF}: The scaling in~\eqref{mom_gen_fun} corresponds to a large deviation asymptotics of the probability
\begin{equation}
P\big(Q_T=j\,\sqrt{T}\big)\sim\mathrm{e}^{-\sqrt{T}\phi_{\text{si}}(j)}\label{eq:LDF asymp}
\end{equation}
where $\phi_{\text{si}}(j)$ is the ldf, related to $\mu_{\text{si}}(\lambda)$ (reference to $\rho_{a(b)}$ is ignored) by a Legendre-Fenchel transformation~\cite{2009_Touchette_The,SM}
$\phi_{\text{si}}(j)=\max_{\lambda}{[j\lambda-\mu_{\text{si}}(\lambda)]}$. Using the transformation, it is immediate that~\eqref{gallavotti_cohen} reflects the symmetry
\begin{equation} \label{gallavotti_cohen_LDF}
\phi_{\text{si}}(j)-\phi_{\text{si}}(-j)=j\log{\frac{\rho_b\,(1-\rho_a)}{\rho_a\,(1-\rho_b)}}.
\end{equation}
Similarly, the asymptotics
$\mu_{\text{si}}(\lambda)\simeq[(4\sqrt{2})/(3\pi)](\log{\omega})^{3/2}$
of~\eqref{musi_complete} for large positive $\lambda$ correspond~\cite{2009_Derrida_Current} to the asymptotics of the ldf
\begin{equation}\label{asymptotes_LDF}
\phi_{\text{si}}(j)\simeq\frac{\pi^2j^3}{24}-j\log{\big(\rho_a\,(1-\rho_b)\big)}
\end{equation}
for large positive $j$. A similar analysis for large negative $j$ gives asymptotics~\eqref{asymptotes_LDF} with $j\to-j$ and $\rho_a\leftrightarrow\rho_b$. Notably, for the non-equilibrium condition $\rho_a\ne\rho_b$, the ldf is skewed.

\textit{Derivation}: In the following, we outline our derivation of~\eqref{musi_complete} within the fluctuating hydrodynamics framework of MFT. The crucial idea behind MFT is to recognize the relevant hydrodynamic modes for a coarse-grained description of the dynamics and characterize the probability of their evolution in terms of an Action, which is analogous to the Martin-Siggia-Rose-Janssen-De Dominicis (MSRJD) Action~\cite{1973_Martin_Statistical,1976_Janssen_On,1978_Dominicis_Dynamics,1978_Dominicis_Field} of the associated fluctuating hydrodynamics equation. For SSEP, the relevant hydrodynamic mode is the locally conserved density $\rho(x,t)$ evolving by~\cite{2023_Saha_Current,SM}
\begin{equation}
\partial_t \rho=\partial_x^2\rho+\frac{1}{T^{1/4}}\,\partial_x\big(\sqrt{\sigma(\rho)}\,\eta\big)
\end{equation}
where $\sigma(\rho)=2\rho(1-\rho)$, and $\eta(x,t)$ is a delta-correlated Gaussian white noise with unit covariance. Corresponding MSRJD-Action on the semi-infinite line is~\cite{2023_Saha_Current,SM}
\begin{equation}\label{eq:S0}
S_0=\mathcal{F}+\int_0^1\mathrm{d}t\int_0^{\infty}\mathrm{d}x\bigg[\widehat{\rho}\,\partial_t\rho-\bigg(\frac{\sigma(\rho)}{2}\partial_x\widehat{\rho}+\partial_x\rho\bigg)\partial_x\widehat{\rho}\bigg]
\end{equation}
where $\widehat{\rho}$ is the response field and $\mathcal{F}$ incorporates contributions from fluctuations in the initial state~\cite{2009_Derrida_Current2,2023_Saha_Current}. Within this description, the generating function $\langle\mathrm{e}^{\lambda Q_T}\rangle=\int \mathcal{D}\left[\rho,\widehat{\rho}\right]\,\mathrm{e}^{-\sqrt{T}S_\lambda\left[\rho,\widehat{\rho}\right]}$ with $S_\lambda=S_0-\lambda Q_T/\sqrt{T}$ and $Q_T$ in~\eqref{eq:QT}.

For large $T$, the path-integral is dominated by a saddle point, leading to~\eqref{mom_gen_fun} with
\begin{equation}
\mu_{\text{si}}(\lambda)\simeq-\min_{\rho,\widehat{\rho}}{S_\lambda\left[\rho,\widehat{\rho}\right]}\equiv-S_\lambda\left[q_{\text{si}},p_{\text{si}}\right] \label{hydro_action}
\end{equation}
(reference to $\rho_{a(b)}$ is ignored) where $(q_{\text{si}},p_{\text{si}})$ is the least-Action path for $(\rho,\widehat{\rho})$. This way, the problem reduces~\cite{2009_Derrida_Current2,2023_Saha_Current} to solving the corresponding Euler-Lagrange equations 
\begin{subequations}\label{euler_lag}
\begin{align}\label{euler_lag_1}
\partial_tq_{\text{si}}&=\partial_x^2q_{\text{si}}-\partial_x[\sigma(q_{\text{si}})\,\partial_xp_{\text{si}}],\quad\text{and}\\
\partial_tp_{\text{si}}&=-\partial_{x}^{2}p_{\text{si}}-\frac{\sigma'(q_{\text{si}})}{2}\,(\partial_xp_{\text{si}})^2\label{euler_lag_2}
\end{align}\end{subequations}
in the semi-infinite domain $x>
0$, with the temporal boundary conditions 
\begin{equation}
p_{\text{si}}(x,0)=\lambda+\int_{\rho_b}^{q_{\text{si}}(x,0)}\frac{2\,\mathrm{d}r}{\sigma(r)}\quad\text{and}\quad p_{\text{si}}(x,1)=\lambda, \label{temp_cond}
\end{equation}
where the integral in $p_{\text{si}}(x,0)$ is the contribution~\cite{2009_Derrida_Current2} from $\mathcal{F}$ for the initial state of SSEP with Bernoulli measure.
Additional spatial boundary conditions,
\begin{equation}\label{semi_spatial_boundary}
q_{\text{si}}(0,t)=\rho_a\quad\text{and}\quad p_{\text{si}}(0,t)=0
\end{equation}
are due to the ``fast-coupling''~\cite{2024_Saha_Large,SM} with reservoir. 

This variational problem is reminiscent of the corresponding problem on the infinite line~\cite{2009_Derrida_Current2} for a Bernoulli-measured initial state with an average density $\rho_a$ for $x<0$ and $\rho_b$ for $x>0$. The only differences between the two problems are the domain $x$ and the boundary conditions. The infinite line problem was formulated in~\cite{2009_Derrida_Current2} within MFT, which was recently solved in~\cite{2022_Mallick_Exact} by identifying an ingenious mapping to the classical integrable system and employing the inverse scattering method. Despite having small differences, the semi-infinite case poses a new nontrivial problem~\cite{1975_Ablowitz_The}, which is incredibly difficult to solve.

A simplification arises for the special choice of initial density pair $(\rho_a,\rho_b)\equiv(1/2,0)$ for the semi-infinite line problem. For this choice of densities, there are no fluctuations in the initial state, which amounts to setting $\mathcal{F}=0$ in~\eqref{eq:S0} with the condition
\begin{equation}
q_{\text{si}}(x,0)=0.\label{eq:q0=0}
\end{equation}
The Euler-Lagrange equations for the corresponding variational problem~\eqref{hydro_action} remains the same as in~\eqref{euler_lag}. The only difference comes in the initial condition in~\eqref{temp_cond}, which is now replaced by~\eqref{eq:q0=0}. This initial condition corresponds to the quenched averaging~\cite{2009_Derrida_Current2,2024_Saha_Large}.

We now show that this quenched semi-infinite line problem has a direct mapping to the quenched infinite line problem with $(\rho_a,\rho_b)\equiv(1,0)$ and fugacity $\tilde{\lambda}=2\lambda$. For the latter problem, the Euler-Lagrange equation is the same~\cite{2009_Derrida_Current2} as in~\eqref{euler_lag}, but now on the entire real line $x$ with the temporal boundary conditions
\begin{equation}\label{infinite_boundary_condition}
q_\text{inf}(x,0)=\theta(-x)\quad\text{and}\quad p_\text{inf}(x,1)=\tilde{\lambda} \,\theta(x)
\end{equation}
(subscript `$\text{inf}$' denotes infinite and $\theta(x)$ is the Heaviside step function).
It is simple to check that the solution admits the symmetry
\begin{subequations}
\begin{align}
q_\text{inf}(x,t) &= 1 - q_\text{inf}(-x, t)\label{Symmetry1}\\
p_\text{inf}(x,t) &= \tilde{\lambda} - p_\text{inf}(-x, t)\label{Symmetry2}
\end{align}
\end{subequations}
which fixes the value of the fields at the origin
\begin{equation}
q_\text{inf}(0,t)=\frac{1}{2}\quad\text{and}\quad p_\text{inf}(0,t)=\frac{\tilde{\lambda}}{2}
\end{equation}
at all times $0<t<T$.
This conclusion is an essential part of our observation, as now, the solution $q_\text{inf}(x,t)$ on positive line $x$ satisfies the boundary conditions (\ref{semi_spatial_boundary},\,\ref{eq:q0=0}) of the semi-infinite line problem with $(\rho_a,\rho_b)\equiv(1/2,0)$. For a similar correspondence of the response field, we define
\begin{equation}
\tilde{p}(x,t)=p_\text{inf}(x,t)-\frac{\tilde{\lambda}}{2}
\end{equation}
which too now replicates the boundary conditions $\tilde{p}(x,1)=\lambda$ and $\tilde{p}(0,t)=0$ of the semi-infinite line problem with $\tilde{\lambda} = 2\lambda$. The fields $(q_{\text{inf}},\tilde{p})$ also satisfy the same Euler-Lagrange equations~\eqref{euler_lag}.

Consequently, the least-Action path $(q_\text{si},p_\text{si})$ for the semi-infinite line problem with $(\rho_a,\rho_b)\equiv(1/2,0)$ and fugacity $\lambda$ is related to the corresponding path $(q_\text{inf},p_\text{inf})$ for the infinite line problem with $(\rho_a,\rho_b)\equiv(1,0)$ and fugacity $2\lambda$ by
\begin{equation}
q_\text{si}(x,t)=q_\text{inf}(x,t)\quad\text{and}\quad p_\text{si}(x,t)=p_\text{inf}(x,t)-\lambda
\end{equation}
for $x\ge 0$ at all times. This correspondence relates the least-Action of the two problems resulting~\cite{SM} in our crucial observation:
\begin{equation}
\mu_{\text{si}}\Big(\lambda,\frac{1}{2},0\Big)=\frac{1}{2}\,\mu_{\text{inf}}(2\lambda,1,0)
\end{equation}
where the pre-factor $1/2$ comes from the half-domain of integration of $x$ in~\eqref{eq:S0}.
The latter scgf is known from the seminal work~\cite{2009_Derrida_Current} of Derrida and Gerschenfeld, which culminates in
\begin{equation}\label{CGF_for_half-zero}
\mu_{\text{si}}\Big(\lambda,\frac{1}{2},0\Big)=\int_{-\infty}^\infty\frac{\mathrm{d}k}{2\pi}\,\log{[1+(\mathrm{e}^{2\lambda}-1)\mathrm{e}^{-k^2}]}.
\end{equation}

The result~\eqref{CGF_for_half-zero} is for the specific initial density pair $(\rho_a,\rho_b)\equiv(1/2,0)$. For extending the result for other densities, we invoke a well-known rotational symmetry~\cite{2009_Derrida_Current2,2010_Lecomte_Current,SM} of the least-Action~\eqref{hydro_action}. Essentially, the least-Action paths for two sets of parameters $(\lambda, \rho_a, \rho_b)$ and $(\lambda', {\rho_a}', {\rho_b}')$ are related under a canonical transformation~\cite{2009_Derrida_Current2,2010_Lecomte_Current,SM}. The symmetry results~\cite{2009_Derrida_Current,2010_Lecomte_Current,SM} in a dependence \eqref{mu_si} of the scgf on $(\lambda,\rho_a,\rho_b)$ through a single parameter $\omega$ defined in~\eqref{ssep_parameter}. This dependence enables us~\cite{SM} to deduce the function $R_{\text{si}}(\omega)$ from the result~\eqref{CGF_for_half-zero}, leading to the expression~\eqref{PW_Rsi}. Note that the $\omega$-dependence does not extend \cite{2004_Derrida_Current,2022_Mallick_Exact} for optimal profile $q_{\textrm{si}}(x,t)$, making it difficult to infer the profile at arbitrary densities.

\textit{Slow bond}: Recent interests~\cite{2017_Baldasso_Exclusion,2019_Franco_Non,2020_Goncalves_Non,2020_Erignoux_HydrodynamicsI,*2020_Erignoux_HydrodynamicsII,2021_Derrida_Large,2024_Saha_Large,2023_Saha_Current} in studying the effects of the coupling strength with reservoir can also be addressed in the semi-infinite line problem. The result~\eqref{musi_complete} is independent~\cite{2023_Saha_Current} of the coupling strength $\gamma$ (see Fig.~\ref{fig: dynamics}) as long as it is larger than $\mathrm{O}(T^{-1/2})$. This is seen from the corresponding result for slow coupling $\gamma=\Gamma/\sqrt{T}$, where the boundary-fluctuations are significant, modifying the scgf~\eqref{mu_si} to
\begin{equation}
\mu_{\text{si}}^{\text{slow}}(\lambda,\rho_a,\rho_b)=\min_z{\Big[\Gamma\sinh^2{(z- u)}+R_{\text{si}}(\sinh^2{z})\Big]} \label{eq:scfg slow si}
\end{equation}
with $\sinh^2{u}=\omega(\lambda,\rho_a,\rho_b)$. In the $\Gamma\to\infty$ limit,~\eqref{mu_si} is recovered from~\eqref{eq:scfg slow si}. The expression~\eqref{eq:scfg slow si} is obtained following an additivity argument~\cite{2021_Derrida_Large}, where contributions in~\eqref{eq:scfg slow si} are separately from the single bond joining the reservoir and the system, and the system itself, optimised over the density at their common site \cite{SM}. This construction is very similar to the discussion in~\cite{2021_Derrida_Large} for a related context, except for a crucial distinction that unlike the latter example, there is no quasi-stationarity for the semi-infinite problem. Despite this, the additivity conjecture gives the correct result~\eqref{eq:scfg slow si}.

A similar additivity argument helps to solve the problem of current fluctuation across a single slow bond at the origin on an infinite one-dimensional lattice~\cite{2013_Franco_Hydrodynamical,2013_Franco_Phase,*2016_Franco_Phase,2020_Erhard_Non}, initially filled with uniform average density $\rho_{a(b)}$ in the negative (positive) half-line. For the hopping rate $\gamma=\Gamma/\sqrt{T}$ across the slow bond, and unity for rest of the lattice, the generating function of current $\big<\mathrm{e}^{\lambda Q_T}\big>\sim\mathrm{e}^{\sqrt{T}\mu_{\text{inf}}^{\text{slow}}(\lambda,\rho_a,\rho_b)}$, for large $T$, with the scgf \cite{SM}
\begin{align}
\mu_{\text{inf}}^{\text{slow}}&(\lambda,\rho_a,\rho_b)=\min_{z_a,z_b}\Big[R_{\text{si}}\big(\sinh^2{z_a}\big)\nonumber\\
&+\Gamma\sinh^2{(z_a+z_b-u)}+R_{\text{si}}\big(\sinh^2{z_b}\big)\Big] \label{eq:scfg slow inf}
\end{align}
where $\sinh^2{u}=\omega(\lambda,\rho_a,\rho_b)$. In the $\Gamma\to\infty$ limit, the celebrated infinite line result~\cite{2009_Derrida_Current,2022_Mallick_Exact} is recovered~\cite{SM} from~\eqref{eq:scfg slow inf}.

\textit{Other applications}: Several other interesting conclusions can be drawn from our result. A known symmetry~\cite{2009_Derrida_Current2} of the MFT-Action extends our exact results to models~\cite{1982_Kipnis_Heat,2007_Giardinà_Duality,2014_Vafayi_Weakly,2022_Franceschini_Symmetric} with a quadratic mobility $\sigma(\rho)=2A\rho(B-\rho)$, where $A$ and $B$ are arbitrary constants, culminating in 
$\mu_{\sigma}(\lambda,\rho_a,\rho_b)=(1/A)\,\mu_{\text{ssep}}\big(\omega(AB\lambda,\rho_a/B,\rho_b/B)\big)
$. Appropriate limits of $A$ and $B$ give the result for the Kipnis-Marchioro-Presutti (KMP) model \cite{1982_Kipnis_Heat,2007_Giardinà_Duality}, the symmetric simple inclusion process (SSIP) \cite{2014_Vafayi_Weakly,2022_Franceschini_Symmetric}, and the non-interacting particles~\cite{2023_Saha_Current}.

The semi-infinite line result provides solutions for two closely related problems. The first problem~\cite{2012_Krapivsky_Symmetric, 2014_Krapivsky_Lattice} concerns an infinite line SSEP with fast injection at a single site, a scenario which serves as the simplest description of hard-core particles spreading on an initially empty system~\cite{2019_Krapivsky_Free,*2020_Krapivsky_Free}. Such dynamics arise in diverse contexts, including the voter model~\cite{2003_Mobilia_vote}, chemical catalysis~\cite{1992_Krapivsky_monomer,1996_Krapivsky_catalytic}, and wetting thin film spreading~\cite{Popescu_2012_wetting_films}. At long-times $t$, the net injection of particles $N_t$ on an empty lattice follows the asymptotics $\langle \mathrm{e}^{\lambda N_t}\rangle\sim\exp{[2\sqrt{t}\mu_{\text{si}}(\lambda,1,0)]}$ with~\eqref{musi_complete}, which results from the statistical independence of the two halves due to fast injection. The second problem concerns the survival probability $P_t(k)$ of the $k^{\text{th}}$ tagged particle (tracer) up to time $t$ in a fully packed SSEP on $\mathbb{Z}^{+}$ in the presence of an absorbing site at the origin. The problem has applications to first passage~\cite{2013_Majumdar_first_passage,2014_Benichou_first_passage} problems, target search~\cite{2011_Benichou_search}, and diffusion limited reactions \cite{rice1985diffusion}. The asymptotics~\eqref{eq:LDF asymp} and the relation $P_t(k)=P(-Q_t>k)$ in the semi-infinite SSEP gives a stretched exponential decay $P_t(k)\sim\exp{(-\sqrt{t/\tau_s})}$ for $k\lesssim2\sqrt{t/\pi}$ at long-times $t$, with $\tau_s=[\phi(-k/\sqrt{t})]^{-2}$.

The final application we discuss is about similar stretched exponential decays in kinetically constrained models~\cite{1983_Skinner_Kinetic,1989_Spohn_Stretched,2020_Gupta_Size,2024_Mukherjee_Stretched}, such as the energy-conserving spin-flip dynamics~\cite{1989_Spohn_Stretched,2024_Mukherjee_Stretched}, where spin auto-correlation decays as $\exp(-\sqrt{t/\tau})$, with $\tau$ remained undetermined. For the spin-model, rigorous result about $\tau$ comes~\cite{SM} from realizing that the domain-wall dynamics is equivalent to the infinite-line SSEP at a uniform density $\rho$. This correspondence relates~\cite{SM} the spin-auto-correlation to the generating function of current in the infinite SSEP leading to $\tau=[\mu_{\text{inf}}(\mathrm{i}\pi,\rho,\rho)]^{-2}$. Similar correspondence holds in presence of a slow bond or in semi-infinite geometry.

\textit{Conclusions}: 
We derived full counting statistics \eqref{musi_complete} for SSEP on a semi-infinite line, which represents an intermediate non-equilibrium regime between the well-studied finite \cite{2004_Bodineau_Current,2005_Bertini_Current} and infinite SSEP \cite{2009_Derrida_Current,2022_Mallick_Exact}. Compared to the latter examples, the scgf \eqref{musi_complete} exhibits a piecewise solution \eqref{PW_Rsi}. Our exact result not only solves the absorbing boundary problem relevant to target survival \cite{2011_Benichou_search}, but also leads to the full counting statistics \eqref{eq:scfg slow inf} for the infinite-SSEP with a defect bond. The variational construction used for the latter solution offers a powerful approach for studying inhomogeneous transport, even for asymmetric dynamics.

There remain several related open problems of immediate interest. Most prominent among them is the scgf for fixed (quenched) initial states. The only available non-trivial result~\cite{2009_Derrida_Current2} is for the infinite line SSEP at half-filling, obtained following a relation with the fluctuating (annealed) initial state.
A similar mapping for the semi-infinite SSEP yields the quenched scgf
\begin{equation}
\mu_{\text{si}}^{\text{que}}\Big(\lambda,\frac{1}{2},\frac{1}{2}\Big)=\int_{-\infty}^\infty\frac{\mathrm{d}k}{\sqrt{8}\,\pi}\log{\big[1+\sinh^2{(\lambda)}\,\mathrm{e}^{-k^2}\big]}.
\end{equation}
An extension of these results in higher dimensions would be interesting, where only limited results are available~\cite{2013_Akkermans_Universal, Becker_2015, Hutardo_2016, Hutardo_2017, Harris_2016}. From a practical point of view, the emergence of the SSEP in quantum circuits~\cite{2023_McCulloch_Full} or hydrodynamics in chaotic quantum systems \cite{2024_Wienand_Emergence} offers exciting applications, where semi-infinite analog \cite{turkeshi2024} becomes relevant. Along similar lines, extensions of our results for quantum analogues of SSEP~\cite{2019_Bernard_Open,2021_Bernard_Can} or for integrable models~\cite{2020_Doyon_Lectures,2022_Bastianello_Introduction} would be timely.

\begin{acknowledgments}
\textit{Acknowledgments}: We acknowledge the financial support of the Department of Atomic Energy, Government of India, under Project Identification No. RTI 4002. TS thanks Bernard Derrida for insightful discussions on the problem. The importance of semi-infinite line SSEP and the additivity ansatz for slow bond originated from those discussions. TS also thanks Mustansir Barma for introducing him to the works on stretched exponential decays, particularly the open problem in~\cite{1989_Spohn_Stretched}. TS
thanks the support from the International
Research Project (IRP) titled `Classical and quantum dynamics in out of equilibrium systems' by CNRS, France.
\end{acknowledgments}

\newpage

\section{End Matter}
\begin{figure}[h]
\centering
\includegraphics[width=\linewidth]{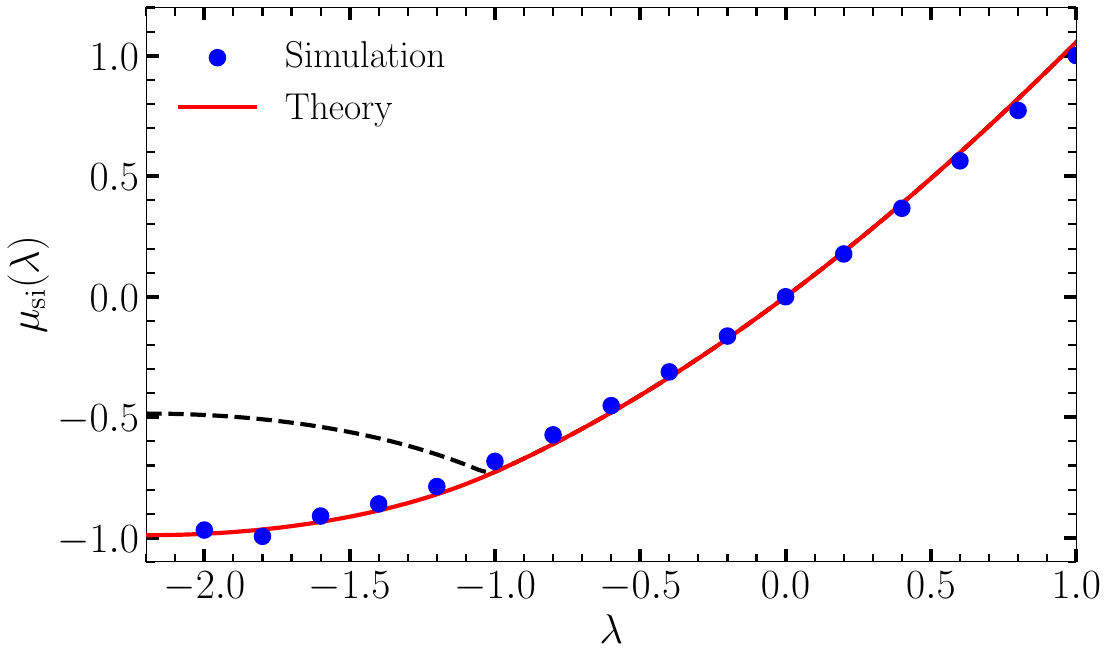}
\caption{\textit{Scgf for semi-infinite SSEP:} The scgf of current for the semi-infinite SSEP for densities $(\rho_a,\rho_b)\equiv(0.9,0.1)$. The solid red line represents the theoretical result \eqref{musi_complete}, while the data points are obtained by the cloning algorithm with $T=500$ and $N_c=50,000$. The dashed line represents $R(\omega(\lambda,\rho_a,\rho_b))$ in \eqref{IntegralRep} as a function of $\lambda$ which differs from the solid line for the parameter regime $\omega(\lambda,\rho_a,\rho_b)<-1/2$, which corresponds to $\lambda\lesssim-1.0$ in the figure.}
\label{fig:pwscgf}
\end{figure}

\textit{The piecewise function}: The scgf \eqref{musi_complete} is plotted in Fig.~\ref{fig:pwscgf} for parameter regimes where the piecewise nature of the function \eqref{PW_Rsi} is relevant. The function $R(\omega)$ in \eqref{IntegralRep} is singular at $\omega=-1/2$, and this leads to a non-convex $R(\omega(\lambda,\rho_a,\rho_b))$ as a function of $\lambda$, indicated by the dashed line in Fig.~\ref{fig:pwscgf}. The piecewise function \eqref{PW_Rsi} constructed from $R(\omega)$ gives an analytic, convex scgf, which agrees with the data from rare event simulation using a cloning algorithm. 

\textit{Cumulants}: The scgf~\eqref{musi_complete} encapsulates all cumulants of $Q_T$ in the large $T$ limit for the semi-infinite SSEP. While the first three cumulants were initially reported in~\cite{2023_Saha_Current}, we present here the fourth cumulant for $\rho_a=\rho_b=\rho$
\begin{equation}
\frac{{\big<Q_T^4\big>}_c}{\sqrt{T}}\simeq\frac{4}{\sqrt{\pi}}\,\rho\,(1-\rho)\Big[1-12\,\big(\sqrt{2}-1\big)\,\rho\,(1-\rho)\Big].
\end{equation}

\textit{Infinite line SSEP with a slow bond}: The scgf of current for an infinite line SSEP with a single slow bond can be obtained by treating the entire system as composed of three subsystems: two semi-infinite systems with unit hopping rates coupled to each other by a single slow bond with hopping rate $\Gamma/\sqrt{T}$.
This leads to a variational formula~\cite{SM}
\begin{align}\label{eq:mu additivity rho lambda inf}
\mu_{\text{inf}}^{\text{slow}}(\lambda,\rho_a,\rho_b)=&\max_{\rho_0,\rho_1}\min_{\lambda_0,\lambda_1}\Big[R_{\text{si}}\big(\omega(\lambda_0,\rho_a,\rho_0)\big)\nonumber\\+\Gamma\,\omega(\lambda_1-\lambda_0,&\rho_0,\rho_1)+R_{\text{si}}\big(\omega(\lambda-\lambda_1,\rho_1,\rho_b)\big)\Big],
\end{align}
where $\rho_{0(1)}$ is the density of the common site between the left (right) semi-infinite lattice and the slow bond. The analysis is similar to the discussion in~\cite{2021_Derrida_Large} for a related problem. 

\begin{figure}[htpt!]
\centering
\includegraphics[width=\linewidth]{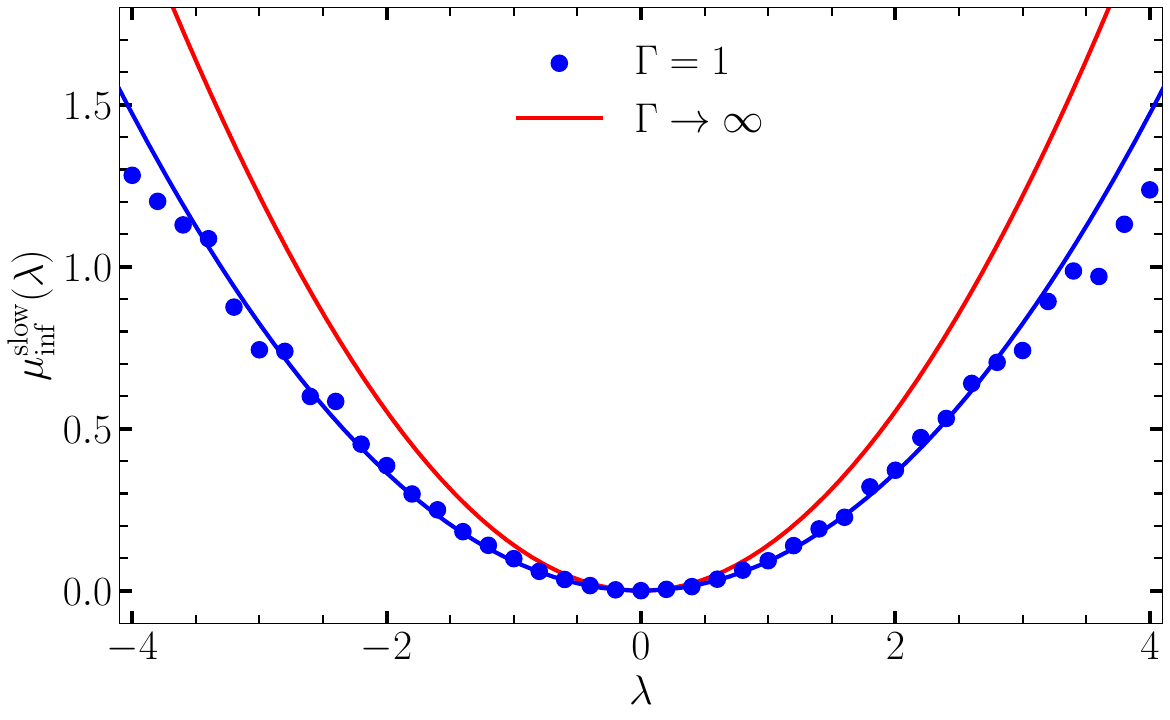}
\caption{\textit{Scgf for infinite SSEP with a slow bond}: The solid blue line denotes the scgf given in \eqref{eq:scfg slow inf} for $\rho_{a(b)}=0.5$ and $\Gamma=1$. The blue markers denote the corresponding numerical results obtained by a continuous-time cloning algorithm with $T=500$ and $N_c=10^4$, for $\Gamma=1$. The solid red line denotes the $\Gamma \to \infty $ limit result~\cite{2009_Derrida_Current}.}
\label{fig: Slow bond infinite}
\end{figure}

\begin{figure}[htpt!]
\centering
\includegraphics[width=\linewidth]{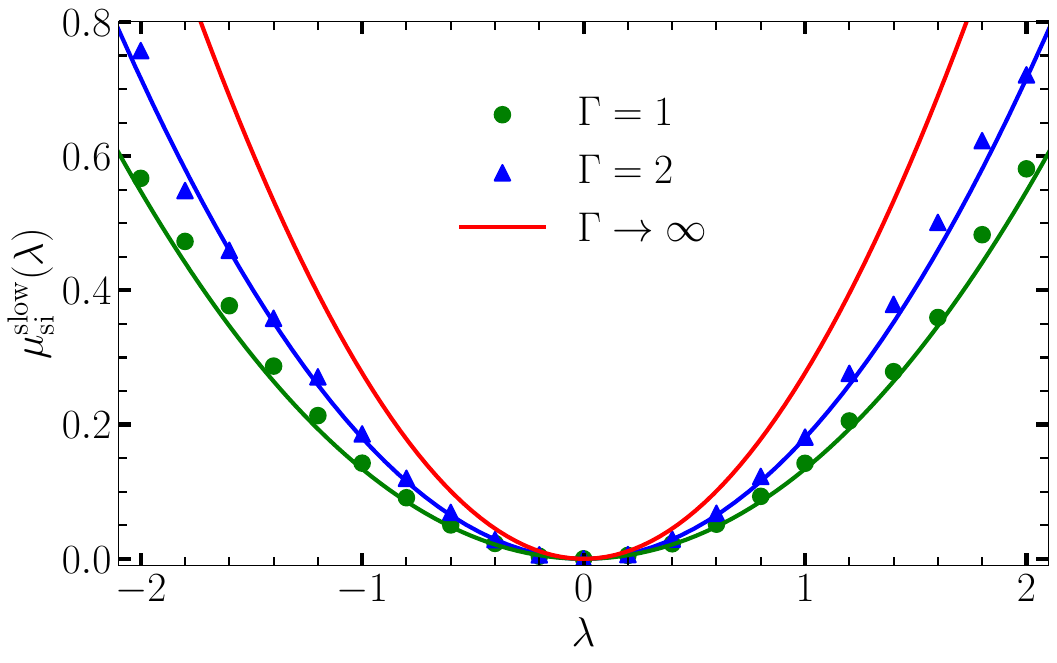}
\caption{\textit{Scgf for semi-infinite SSEP with slow boundary coupling}: The solid lines denote the scgf~\eqref{eq:scfg slow si} for $\rho_{a(b)}=0.5$ and different values of $\Gamma$ (green for $\Gamma=1$, blue for $\Gamma=2$, and red for $\Gamma\to \infty$). The markers denote the corresponding numerical results obtained by the cloning algorithm with $T=500$ and $N_c=10^4$.}
\label{fig: Numerical_slow_bond}
\end{figure}

The expression~\eqref{eq:mu additivity rho lambda inf} further simplifies using Eq.~(29) of~\cite{2021_Derrida_Large} leading to the expression in~\eqref{eq:scfg slow inf}. We confirm this theoretical result by comparing with numerical data from our rare event simulation as shown in Fig.~\ref{fig: Slow bond infinite}.

A similar construction leads to the scgf~\eqref{eq:scfg slow si} for semi-infinite SSEP with slow coupling to reservoir. A comparison of the result with numerical data is shown in Fig.~\ref{fig: Numerical_slow_bond}.

\begin{figure}[htpt!]
\centering
\includegraphics[width=0.95\linewidth]{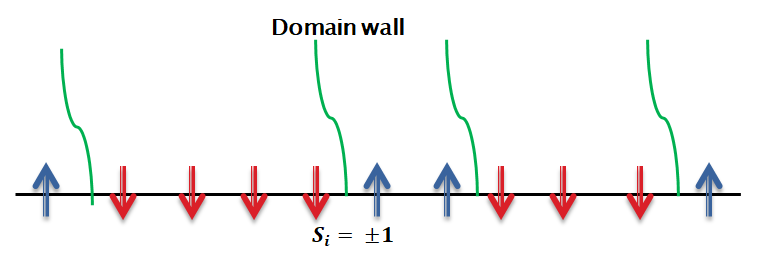}
\caption{Dynamics of spins~\cite{1989_Spohn_Stretched} following a kinetic constraint of fixed energy: a spin can flip with unit rate only if its nearest neighbor's spins are of opposite sign. For example, the fifth spin from left has left neighbor down and right neighbor up, so it is allowed to flip, making the domain wall move leftward by one lattice unit. The movement of domain walls (indicated by green lines) is analogous to the dynamics of particles in a SSEP.}
\label{fig: domain_wall}
\end{figure}

\textit{A connection between SSEP and a kinetically constrained model}: The energy-conserving spin-flip dynamics on a one-dimensional infinite lattice, defined in~\cite{1989_Spohn_Stretched}, where each site $i$ at a given time $t$ is assigned with an Ising spin $S_i(t) = \pm 1$. The dynamics is such that a spin can flip only if its two nearest neighbor's spins have opposite signs, therefore conserving the total energy (Fig.~\ref{fig: domain_wall}). This constraint ensures the conservation of number of domain walls, whose dynamics precisely mirror the particle movement in the SSEP on an infinite lattice.

This correspondence relates the spin-auto-correlation to the particle current in SSEP. We note that whenever a domain wall passes through a site $i$ (see Fig.~\ref{fig: domain_wall}), the associated Ising spin, $S_i$, changes its sign. If $Q_t$ is the net rightward flow of domain walls through the site $i$ in time $t$, then the quantity $S_i(0) S_i(t)$ takes values $+1$ or $-1$ for $Q_t$ being even or odd. This implies the auto-correlation
\begin{equation}\label{correlation}
\big<S_i(0)\,S_i(t)\big>=P_t(\text{even }Q_t) - P_t(\text{odd }Q_t),
\end{equation} 
with the probability $P_t(\text{even }Q_t )$ of $Q_t$ being even and $P_t(\text{odd }Q_t )$ for odd, respectively. 
The difference of the two probabilities in~\eqref{correlation} relates to the generating function of current $\big<\mathrm{e}^{\lambda Q_t}\big>$ in SSEP. Noting that $Q_t$ takes only integer values, we see that for $\lambda=\mathrm{i}\pi$,
\begin{equation}\label{current}
\big<\mathrm{e}^{\mathrm{i}\pi Q_t}\big>=\sum_{Q_t}P(Q_t)\,\mathrm{e}^{\mathrm{i}\pi Q_t} =P_t(\text{even }Q_t)-P_t(\text{odd }Q_t).
\end{equation}
From~\eqref{correlation} and~\eqref{current}, it is evident that at large times, the spin-auto-correlation follows the asymptotics of the current statistics in SSEP. Using the results for the latter from~\cite{2009_Derrida_Current} we see that for large times $t$, 
\begin{equation}\label{spin asymptotics}
\big<S_i(0)\,S_i(t)\big>\sim\mathrm{e}^{\sqrt{t}\mu_{\text{inf}}(\mathrm{i}\pi,\rho,\rho)}=\mathrm{e}^{-\sqrt{{t}/{\tau}}}
\end{equation}
with $\tau=[\mu_{\text{inf}}(\mathrm{i}\pi,\rho,\rho)]^{-2}$. Here, the density $\rho$ is the average uniform density of domain walls in the spin model~\cite{1989_Spohn_Stretched}. The exact result~\eqref{spin asymptotics} is consistent with the bounds derived in~\cite{1989_Spohn_Stretched}.

\bibliography{letter}

\end{document}


\title{Supplementary Material: Large deviations of current for the symmetric simple exclusion process on a semi-infinite line and on an infinite line with a slow bond}

\author{Kapil Sharma}
\email{kapil.sharma@tifr.res.in}
\affiliation{Department of Theoretical Physics, Tata Institute of Fundamental Research, Homi Bhabha Road, Mumbai 400005, India}

\author{Soumyabrata Saha}
\email{soumyabrata.saha@tifr.res.in}
\affiliation{Department of Theoretical Physics, Tata Institute of Fundamental Research, Homi Bhabha Road, Mumbai 400005, India}

\author{Sandeep Jangid}
\email{sandeep.jangid@tifr.res.in}
\affiliation{Department of Theoretical Physics, Tata Institute of Fundamental Research, Homi Bhabha Road, Mumbai 400005, India}

\author{Tridib Sadhu}
\email{tridib@theory.tifr.res.in}
\affiliation{Department of Theoretical Physics, Tata Institute of Fundamental Research, Homi Bhabha Road, Mumbai 400005, India}

\date{\today}

\begin{abstract}
In this Supplement, we present additional details complementing the \textit{Letter}. We discuss details about our numerical simulation providing results on finite-size effects, optimal density profile at the final time as well as scgf for a non-equilibrium initial condition. Subsequently, we explicitly write the MFT-Action for the current generating function of the SSEP on the semi-infinite line with slow coupling to the reservoir. Additional details about a symmetry of the least-Action leading to the simplified parameter dependence of the scgf are discussed. Certain steps of our calculations quoted in the \textit{Letter} are presented in the later parts including the semi-infinite line with slow coupling to a reservoir.
\end{abstract}

\maketitle

\tableofcontents

\section{Numerical details: The cloning algorithm and more results}
\begin{figure}[t]
\centering
\includegraphics[width=0.99\linewidth]{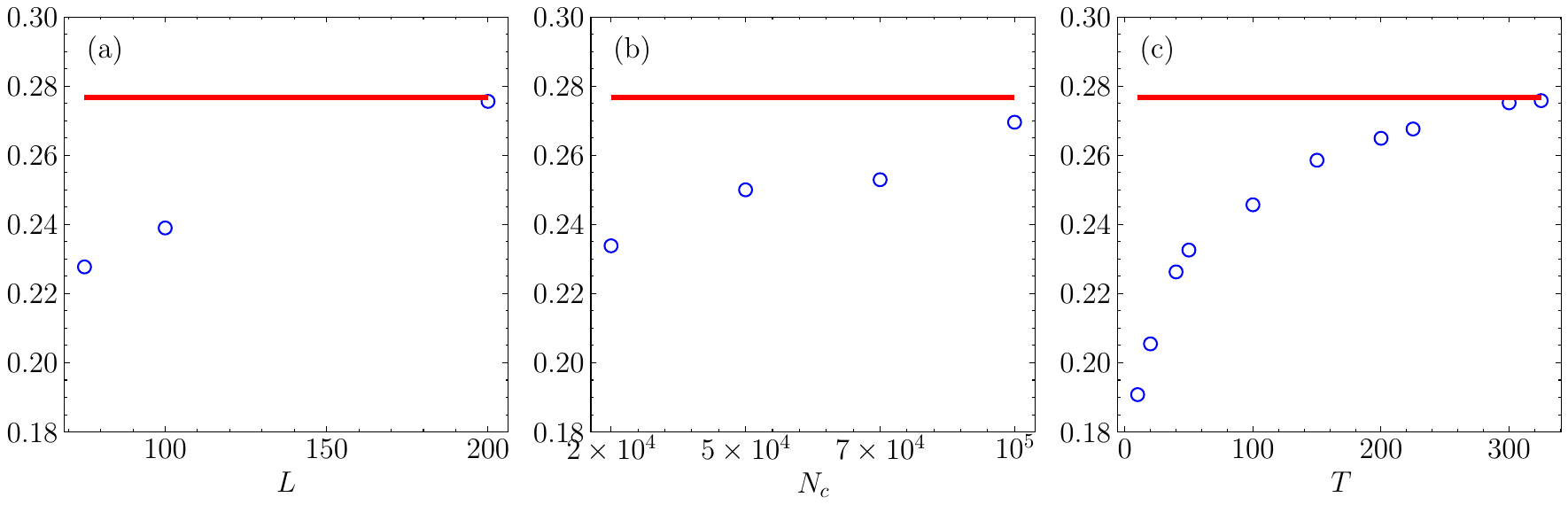} \caption{The solid red line represents the theoretical value of the scgf $\mu_{\text{si}}(\lambda,\rho_a,\rho_b)$ (eq.~(3) of the \emph{Letter}) for the semi-infinite line SSEP with $\lambda = 1.0$ and $\rho_a=\rho_b=1/2$. Results in (a) are for clone size $N_c = 5 \times 10^{4}$ and time $T = 300$, showing dependence on system length $L$. Results in (b) are for $T = 250$ and $L= 200$, showing dependence on the clone size $N_c$. Results in (c) are for $L = 200$ and $N_c = 10^{5}$, showing convergence to theoretical result at large $T$.}
\label{fig: Supp_finite_size}
\end{figure}

Our numerical results in Fig.~2 of the \emph{Letter} are based on a continuous time cloning algorithm. The cloning algorithm for discrete-time Markov processes was introduced in~\cite{2006_Giardinà_Direct}, and subsequently extended for continuous-time processes in~\cite{2007_Lecomte_Numerical}. We suitably adapted the algorithm in~\cite{2007_Lecomte_Numerical,2019_Perez_Sampling} for obtaining the scgf of current in the semi-infinite line SSEP. Our simulation was done on a finite lattice of $L$ sites, initially populated with average density $\rho_b$, with a reflecting boundary at site $L$, and a reservoir of density $\rho_a$ connected at site $1$ (See Fig.~1 of the \emph{Letter}). The net current $Q_T$ is measured at the left boundary over a period $T\ll L^2$ such that the effect of the reflecting boundary is negligible. The resulting scgf for densities $(\rho_a,\rho_b)\equiv (\frac{1}{2},\frac{1}{2})$, on finite lattice of length $L=200$ at measurement time $T=300$ is shown in Fig.~2 of the \emph{Letter}. The finite size effects due to the reflecting boundary, the measurement time, and the clone size are shown in Fig.~\ref{fig: Supp_finite_size}. Further improvements of numerical result require advanced computational resources, which are not currently available to us.

\begin{figure}[t]
\centering
\includegraphics[width=0.95\linewidth]{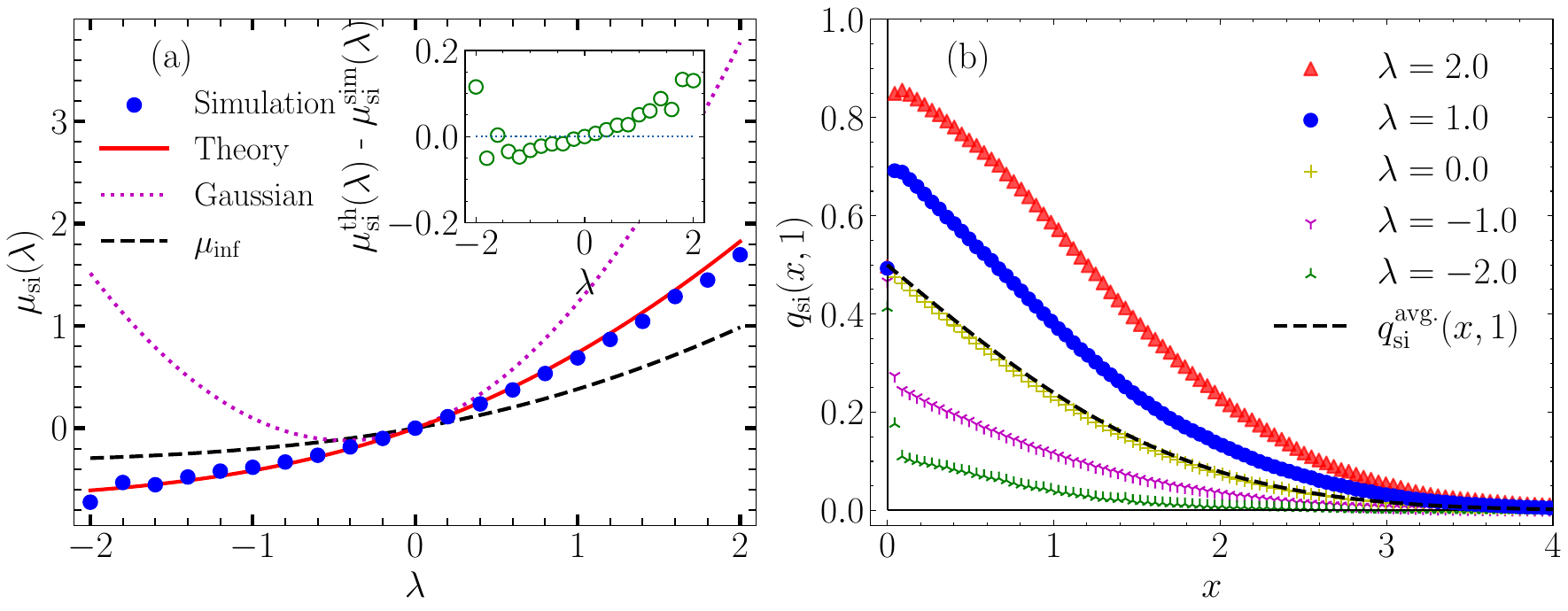}
\caption{(a) Scgf for semi-infinite SSEP with fast boundary coupling in a non-equilibrium initial state. The solid red line represents the theoretical result of scgf (eq.~(3) of the \emph{Letter}) with blue dots representing the corresponding simulation result. There is a good agreement with the deviations shown in the inset. For comparison, gaussian approximation and the corresponding scgf for infinite line $\mu_{\text{inf}}(\lambda)$ are shown in magenta dotted line and black dashed line, respectively.  (b) Simulation result for the \textit{optimal profile} $q_{\textrm{si}}(x,1)$ at the final time plotted against $x = i/\sqrt{T}$, where $i$ is the lattice site index. The markers represent the profile for different values of $\lambda$, indicated in the plot legends. The profile for $\lambda=0$ is the average density profile $q_{\text{si}}^{\text{avg.}}(x, 1)$ and it compares well with the theoretical result given in \cite{2023_Saha_Current}, indicated in black dashed line.  Numerical simulation is done by a continuous-time cloning algorithm for the initial densities $(\rho_a, \rho_b) \equiv (\frac{1}{2}, 0)$ with $T = 400$ and $N_c = 10^{5}$.}
\label{fig: Optimal Profiles and sgf}
\end{figure}

The scgf in Fig.~2 of the \emph{Letter} is for an equilibrium initial state. In Fig~\ref{fig: Optimal Profiles and sgf}(a) we have plotted the corresponding results for a non-equilibrium initial state $(\rho_a, \rho_b)\equiv(\frac{1}{2},0)$. The cloning algorithm also offers the optimal density profile $q_{\text{si}}(x,1)$ at the final time, which is the density averaged over all clones at the final measurement time for a given $\lambda$. The optimal profile is shown in the Fig~\ref{fig: Optimal Profiles and sgf}(b) for different values of $\lambda$ for the densities $(\rho_a, \rho_b)\equiv(\frac{1}{2},0)$.

\section{An Action formulation for the semi-infinite line SSEP coupled to a reservoir}
Here, we present the explicit expression of the MFT-Action for the semi-infinite line SSEP in presence of slow coupling $\gamma=\Gamma/\sqrt{T}$ with the reservoir (see model definition in Fig.~1 in \emph{Letter}). The generating function of the time-integrated current $Q_T$ across the boundary reservoir is given as a path integral~\cite{2023_Saha_Current}
\begin{equation}
\big<\mathrm{e}^{\lambda Q_T}\big>=\int\mathcal{D}\left[\rho,\widehat{\rho}\right]\,\mathrm{e}^{-\sqrt{T}\,S_\lambda[\rho,\widehat{\rho}]} \label{mgf_suppl}
\end{equation}
where the Action is
\begin{equation}
S_\lambda\left[\rho,\widehat{\rho}\right]=-\lambda \int_0^{\infty}\mathrm{d}x\,[\rho(x,1)-\rho(x,0)]+\mathcal{F}[\rho(x,0)]+\int_0^1\mathrm{d}t\,\bigg[\int_0^{\infty}\mathrm{d}x\,\big(\widehat{\rho}(x,t)\,\partial_t\rho(x,t)\big)-H\left[\rho,\widehat{\rho}\right]\bigg]. \label{action_suppl}
\end{equation}
Here, the first term is due to $Q_T$ in eq.~(1) of the \emph{Letter}, and
\begin{equation}
\mathcal{F}\left[\rho\right]=\int_0^\infty\mathrm{d}x\int_{\rho_b}^{\rho}\mathrm{d}r\,\frac{2\,\big(\rho-r\big)}{\sigma(r)} \label{initial_fluct_suppl}
\end{equation}
with $\sigma(\rho)=2\,\rho\,(1-\rho)$, is the contribution~\cite{2009_Derrida_Current2} from the probability of $\rho(x,0)$ in the initial state of the semi-infinite line SSEP populated with Bernoulli measure of average density $\rho_b$. The effective Hamiltonian
\begin{equation}
H\left[\rho,\widehat{\rho}\right]=H_{\text{bdry}}\left[\rho(0,t),\widehat{\rho}(0,t)\right]+\int_0^{\infty}\mathrm{d}x\,\bigg[\frac{\sigma(\rho)}{2}\,\partial_x\widehat{\rho}(x,t)-\partial_x\rho(x,t)\bigg]\,\partial_x\widehat{\rho}(x,t) \label{tot_hamilt_suppl}
\end{equation}
where 
\begin{equation}
H_{\text{bdry}}\left[\rho,\widehat{\rho}\right]=\Gamma\,\big[(\mathrm{e}^{\widehat{\rho}}-1)\,\rho_a\,(1-\rho)+(\mathrm{e}^{-\widehat{\rho}}-1)\,\rho\,(1-\rho_a)\big] \label{bdry_hamilt_suppl}
\end{equation}
is the contribution~\cite{2023_Saha_Current} from fluctuations at the left boundary of the lattice. In~\eqref{action_suppl} there is no condition imposed for the hydrodynamic fields at the left boundary. The boundary condition eq.~(14) of the \emph{Letter} emerges in the fast-coupling $\Gamma\to \infty$ limit, where any small boundary fluctuations bear excessive cost, leading to $\widehat{\rho}(0,t)=0$ for which $H_{\text{bdry}}=0$. This is the MFT-formulation in eq.~(10) of the \emph{Letter} with $\lambda$ set to $0$.

\begin{figure}
\centering
\includegraphics[width=0.6\linewidth]{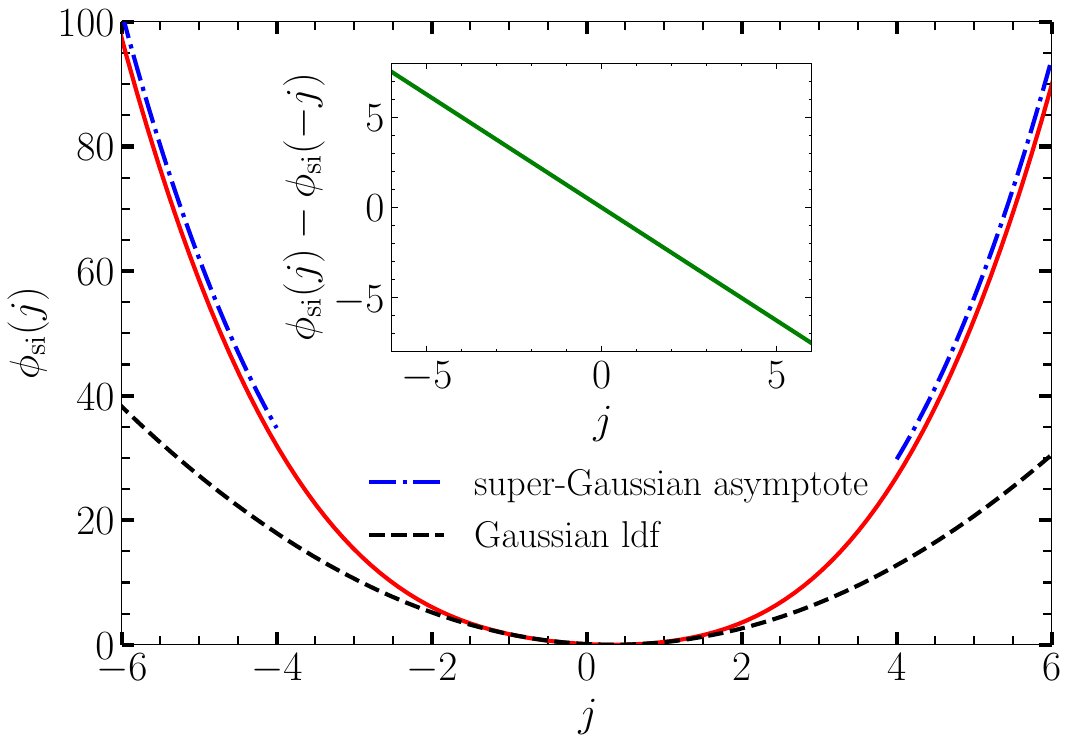}
\caption{The large deviation function indicated in solid red line for $\rho_a=0.6$ and $\rho_b=0.3$. The super-Gaussian asymptotes in eq.~(8) of the \emph{Letter} and the Gaussian ldf with the appropriate mean and variance are indicated in blue dot-dashed and black dashed lines respectively. The inset confirms the fluctuation symmetry in eq~(7) of the \emph{Letter}.}
\label{fig:ldf}
\end{figure}

The variational solution of the MFT in eq.~(12) of the \emph{Letter} gives the scgf for current in eq~(3) of the \emph{Letter}. Corresponding large deviation function generated by numerically evaluating the Legendre-Fenchel transformation of the scgf is shown in Fig.~\ref{fig:ldf}. The non-Gaussian asymptotic behaviour predicted in eq.~(8) of the \emph{Letter} is confirmed in the plot as well as the Gallavotti-Cohen-type fluctuation symmetry relation in eq~(7) of the \emph{Letter} is verified in the inset.

\textit{Remark:} Interestingly, an alternative representation of eq~(3c) of the \emph{Letter},
\begin{equation}
\mu_{\text{si}}(\lambda,\rho_a,\rho_b)=\frac{1}{2\sqrt{\pi}}\sum_{n=1}^\infty\frac{(-1)^{n+1}}{n^{3/2}}\,[4\,\omega\,(1+\omega)]^n \label{scgf_sum_form}
\end{equation}
bears a striking resemblance to the scgf for the number of surviving particles in an assembly of annihilating random walkers \cite{2001_Schutz_Exactly}. This similarity of the two results was brought to our attention by Gunter M. Schütz, and merits further understanding.

\section{Proof of eq.~(21) using eq.~(20) of the \emph{Letter}}
In the fast coupling limit $\Gamma\to \infty$, minimization of the Action~\eqref{action_suppl} corresponds to the optimal profiles $(\rho, \hat{\rho})\equiv(q_{\text{si}},p_{\text{si}})$ having the Euler-Lagrange equations and boundary conditions mentioned in eq.~(13) and eq.~(15) of the $Letter$. Substituting $\partial_tq(x,t)$ from eq.~(12a) of the \emph{Letter} in the expression of $S_\lambda\left[q_{\text{si}},p_{\text{si}}\right]$ from~\eqref{action_suppl}, and subsequently using the spatial boundary conditions of $q_{\text{si}}$ and $p_{\text{si}}$ after performing an integration by parts, the least-Action for the semi-infinite line SSEP with fugacity $\lambda$ and $(\rho_a,\rho_b)\equiv(1/2,0)$ becomes~\cite{2023_Saha_Current}
\begin{equation}\label{action_2}
\mu_{\text{si}}\Big({\lambda},\frac{1}{2},0\Big)=\lambda\int_0^\infty\mathrm{d}x\,\big[q_{\text{si}}(x,1)-q_{\text{si}}(x,0)\big]-\int_0^1\mathrm{d}t\int_0^\infty\mathrm{d}x\,\big[q_{\text{si}}\,(1-q_{\text{si}})\,(\partial_x p_{\text{si}})^2\big],
\end{equation}
while the corresponding least-Action for the infinite line SSEP with fugacity $2\lambda$ and $(\rho_a,\rho_b)\equiv(1,0)$ in terms of the least-Action path is~\cite{2009_Derrida_Current2}
\begin{equation}\label{action_1}
\mu_{\text{inf}}\big(2\lambda, 1, 0\big)=2\lambda\int_{0}^\infty\mathrm{d}x\,\big[q_{\text{inf}}(x,1)-q_{\text{inf}}(x,0)\big]-\int_0^1\mathrm{d}t\int_{-\infty}^\infty\mathrm{d}x\,\big[q_{\text{inf}}\,(1-q_{\text{inf}})\,(\partial_xp_{\text{inf}})^2\big].
\end{equation}
The primary differences between the two expressions (\ref{action_2},~\ref{action_1}) are in the fugacity and the domain of $x$-integration. 

For the density pair $(\rho_a,\rho_b)\equiv(1,0)$ in the infinite lattice problem, using the parity symmetry
\begin{equation}
q_{\text{inf}}(x ,t)=1-q_{\text{inf}}(-x, t)\quad\text{and}\quad p_{\text{inf}}(x, t) =2\lambda-p_{\text{inf}}(-x, t)
\end{equation}
of the least-Action path, the generating function in~\eqref{action_1} reduces to
\begin{equation}\label{action_3}
\mu_{\text{inf}}\big(2\lambda, 1, 0\big)=2\,\Bigg\{\lambda\int_{0}^\infty\mathrm{d}x\,\big[q_{\text{inf}}(x,1)-q_{\text{inf}}(x,0)\big]-\int_0^1\mathrm{d}t\int_{0}^\infty\mathrm{d}x\,\big[q_{\text{inf}}\,(1-q_{\text{inf}})\,(\partial_xp_{\text{inf}})^2\big]\Bigg\}
\end{equation}

The expression in~\eqref{action_3} is similar to~\eqref{action_2}, and subsequently, using the relations
\begin{equation}
q_{\text{si}}(x,t)=q_\text{inf}(x,t)\quad\text{and}\quad p_{\text{si}}(x,t)=p_\text{inf}(x,t)-\lambda
\end{equation}
from eq.~(20) of the \emph{Letter}, it becomes apparent that the two least-Actions (\ref{action_2},~\ref{action_3}) are related by
\begin{equation}
\mu_{\text{si}}\Big(\lambda,\frac{1}{2},0\Big)=\frac{1}{2}\,\mu_{\text{inf}}(2\lambda, 1, 0),
\end{equation}
which is the eq.~(21) of the \emph{Letter}.

\section{Proof of the $\omega$-dependence of the scgf using a rotational symmetry of the Action}
The discussion in this Section follows from~\cite{2009_Derrida_Current2,2010_Lecomte_Current}. The least-Action in eq.~(11) of the \emph{Letter} for the annealed case (fluctuating initial state) has an underlying rotational symmetry that follows from a direct correspondence to the Heisenberg spin chain~\cite{2009_Derrida_Current2,2007_Tailleur_Mapping,*2008_Tailleur_Mapping}. As a consequence of the symmetry, the least-Action corresponding to parameters $(\lambda,\rho_a,\rho_b)$ and $(\lambda',\rho_a',\rho_b')$ are the same as long as the parameters are related by $\omega(\lambda, \rho_a, \rho_b) = \omega(\lambda', \rho'_a, \rho'_b)$, with the function $\omega$ defined in eq.~(3d) of the \emph{Letter}. To see this explicitly, consider the mapping to $\rho'_a = \rho'_b =1/2$ with the corresponding $\lambda'$ related by
\begin{equation}\label{half_half_omega}
\omega(\lambda,\rho_a,\rho_b)=\omega\Big(\lambda',\frac{1}{2},\frac{1}{2}\Big)=\Big(\sinh\frac{\lambda'}{2}\Big)^2.
\end{equation}
A re-parametrization of $\rho_a$ and $\rho_b$ in terms of $u$ and $v$, following
\begin{equation}
\rho_a=\frac{\mathrm{e}^v\cosh{u}-1}{\mathrm{e}^\lambda-1}\quad\text{and}\quad\rho_b=\frac{\mathrm{e}^{-v}\cosh{u}-1}{\mathrm{e}^{-\lambda}-1}
\end{equation}
gives a simple solution $\lambda'=2u$ of~\eqref{half_half_omega}.

In terms of these re-parametrized variables, the least-Action path $(q_{\text{si}},p_{\text{si}})$ in eq.~(11) of the \emph{Letter} corresponding to parameters $(\lambda, \rho_a, \rho_b)$ is related to the least-Action path $(q'_{\text{si}},p'_{\text{si}})$ of the same problem, but corresponding to $(\lambda',1/2,1/2)$ by the transformation~\cite{2009_Derrida_Current2}
\begin{subequations}
\begin{align}
q_{\text{si}}&=\frac{1}{\sinh{u}\sinh{\frac{\lambda}{2}}}\,\bigg[\mathrm{e}^{p_{\text{si}}'-u}\sinh{\frac{\lambda+u-v}{2}}-\sinh{\frac{\lambda-u-v}{2}}\bigg]\,\bigg[q_{\text{si}}'\,\mathrm{e}^{u-p_{\text{si}}'}\sinh{\frac{u+v}{2}}-(1-q_{\text{si}}')\sinh{\frac{u-v}{2}}\bigg],\label{q_transform}\\
p_{\text{si}}&=\log{\bigg[1+\frac{\mathrm{e}^u\,(\mathrm{e}^\lambda-1)\,(\mathrm{e}^{p_{\text{si}}'}-1)}{\mathrm{e}^{p_{\text{si}}'}\,(\mathrm{e}^u-\mathrm{e}^v)+\mathrm{e}^u\,(\mathrm{e}^{u+v}-1)}\bigg]}.\label{p_transform}
\end{align}
\end{subequations}

One way to see this is by noting that the transformed fields (\ref{q_transform},~\ref{p_transform}) keep the bulk Hamiltonian $H\left[q,p\right]$~\eqref{tot_hamilt_suppl} invariant, and thus satisfy the same Euler-Lagrange equations. This has already been shown in~\cite{2009_Derrida_Current2} for the infinite-line problem and the analysis for the semi-infinite-line problem is similar. The important differences between the two problems are in the boundary conditions. For the semi-infinite-line problem, we explicitly verify that the transformation preserves the corresponding structure of the temporal boundary conditions (eq.~(13) of the \emph{Letter}).
\begin{subequations}
\begin{align}
p_{\text{si}}'(x,0)=\lambda'+\int_{1/2}^{q_{\text{si}}'(x,0)}\,\frac{\mathrm{d}r}{r\,(1-r)}\;&\Longrightarrow\;p_{\text{si}}(x,0)=\lambda+\int_{\rho_b}^{q_{\text{si}}(x,0)}\,\frac{\mathrm{d}r}{r\,(1-r)}\;,\quad\text{and}\label{temp_1}\\
p_{\text{si}}'(x,1)=\lambda'\;&\Longrightarrow\;p_{\text{si}}(x,1)=\lambda\label{temp_2}
\end{align}
as well as the spatial boundary conditions
\begin{align}
q_{\text{si}}'(0,t)=\frac{1}{2}\;,\;p_{\text{si}}'(0,t)=0\;&\Longrightarrow\;q_{\text{si}}(0,t)=\rho_a\;,\;p_{\text{si}}(0,t)=0\;,\quad\text{and}\label{spa_1}\\
q_{\text{si}}'(\infty,t)=\frac{1}{2}\;,\;p_{\text{si}}'(\infty,t)=\lambda'\;&\Longrightarrow\;q_{\text{si}}(\infty,t)=\rho_b\;,\;p_{\text{si}}(\infty,t)=\lambda.\label{spa_2}
\end{align}
\label{bound_cond}
\end{subequations}
The invariance of the Euler-Lagrange equations along with~\eqref{bound_cond} culminate into 
\begin{equation}\label{eq:mu mu reln}
\mu(\lambda,\rho_a,\rho_b)=\mu\Big(\lambda',\frac{1}{2},\frac{1}{2}\Big).
\end{equation}
Using~\eqref{half_half_omega} to express $\lambda'$ in terms of $\omega(\lambda,\rho_a,\rho_b)$ in~\eqref{eq:mu mu reln}, we arrive at
\begin{equation}
\mu(\lambda, \rho_a, \rho_b) = \mu\Big(2\,\text{arcsinh}\,\sqrt{\omega(\lambda,\rho_a,\rho_b)}, \frac{1}{2}, \frac{1}{2}\Big),
\end{equation}
thus proving the $\omega$ dependence of the scgf for the semi-infinite line SSEP.

\section{Semi-infinite line SSEP \emph{slowly} coupled to a reservoir}
For slow coupling $\gamma=\Gamma/\sqrt{T}$ with the reservoir, contributions from the fluctuations at the boundary are relevant, as evidenced in the Hamiltonian~\eqref{tot_hamilt_suppl}. Effect of these boundary fluctuations can be incorporated in the current fluctuations using the additivity principle~\cite{2021_Derrida_Large}, whose basic premise is to treat the entire system as composed of two subsystems: the bond linking the reservoir with the semi-infinite lattice, and the lattice itself. The main idea is then to assume that for current fluctuations at large times, the two subsystems are independent of each other, except their dependence through the density at their common point, which is adjusted to maximize the probability of current 
\begin{equation}\label{eq:additivity prob equation}
P(Q_T)\simeq\max_{\rho_0}P_{\text{bond}}(Q_T,\rho_a,\rho_0)\,P_{\text{si}}(Q_T,\rho_0,\rho_b)
\end{equation}
for large $T$. In this description, the single slow bond is coupled at the two ends with reservoirs $\rho_a$ and $\rho_0$, and the semi-infinite lattice is coupled to a reservoir of density $\rho_0$ at its left end. For long times $T$, $P_{\text{bond}}(Q_T=j\sqrt{T})\sim\mathrm{e}^{-\sqrt{T}\phi_{\text{bond}}(j,\rho_a,\rho_0)}$, where $\phi_{\text{bond}}(j,\rho_a,\rho_0)$ is the Legendre-Fenchel transformation of $\Gamma\,\omega(\lambda,\rho_a,\rho_0)$~\cite{2021_Derrida_Large} with the function $\omega$ defined in eq.~(3d) of the \emph{Letter}. Using similar asymptotics (eq.~(6) of the \emph{Letter}) for the semi-infinite lattice, 
\eqref{eq:additivity prob equation} gives the large deviation asymptotics $P(Q_T=j\sqrt{T})\sim\mathrm{e}^{-\sqrt{T}\phi_{\text{si}}^{\text{slow}}(j,\rho_a,\rho_b)}$ with the ldf
\begin{equation}\label{eq:additivity for ldf}
\phi_{\text{si}}^{\text{slow}}(j,\rho_a,\rho_b)=\min_{\rho_0}\big[\phi_{\text{bond}}(j,\rho_a,\rho_0)+\phi_{\text{si}}(j,\rho_a,\rho_b)\big].
\end{equation}

For the corresponding Legendre-Fenchel transformations,~\eqref{eq:additivity for ldf} gives~\cite{2021_Derrida_Large}
\begin{equation}\label{eq:additivity in scgf}
\mu_{\text{si}}^{\text{slow}}(\lambda,\rho_a,\rho_b)=\max_{\rho(0)}\min_{\lambda_0}\Big[\Gamma\,\omega(\lambda_0,\rho_a,\rho(0))+R_{\text{si}}\big(\omega(\lambda-\lambda_0,\rho(0),\rho_b)\big)\Big],
\end{equation}
where $R_{\text{si}}$ is defined in eq.~(3c) of the \emph{Letter}.

The expression~\eqref{eq:additivity in scgf} further simplifies using an identity (see eq.~(29) of~\cite{2021_Derrida_Large}), which leads to
\begin{equation}\label{eq:mu z variational}
\mu_{\text{si}}^{\text{slow}}(\lambda,\rho_a,\rho_b)=\min_{z}\Big[\Gamma\sinh^2{(z-u)}+R_{\text{si}}\big(\sinh^2{z}\big)\Big],
\end{equation}
with $\sinh^2{u}=\omega(\lambda,\rho_a,\rho_b)$ as reported in eq.~(23) of the \emph{Letter}. 

The variational problem in~\eqref{eq:mu z variational} could be explicitly solved using the expression for $R_{\text{si}}$ in eq.~(3c) of the \emph{Letter}, leading to a parametric expression
\begin{subequations}
\begin{equation}
\mu_{\text{si}}^{\text{slow}}(\lambda,\rho_a,\rho_b)=\Gamma\sinh^2{(z-u)}+\int_{-\infty}^\infty\frac{\mathrm{d}k}{2\pi}\log{\big[1+\sinh^2{(2z)}\,\mathrm{e}^{-k^2}\big]}
\end{equation}
where $z$ is given by the solution of the equation
\begin{equation}
\sinh{2(u-z)}=\frac{\sinh{4z}}{\Gamma\pi}\int_{-\infty}^\infty\frac{\mathrm{d}k}{\sinh^2{2z}+\mathrm{e}^{k^2}}
\end{equation}
\end{subequations}
with $\sinh^2{u}=\omega(\lambda,\rho_a,\rho_b)$.

\section{Recovering the `fast' infinite line result from eq. (24) of the \emph{Letter}}
A self-consistency check of the scgf of current for the infinite line SSEP with one slow bond is to recover the well-known result~\cite{2009_Derrida_Current,2022_Mallick_Exact,*2024_Mallick_Exact} for the canonical problem in the fast bond limit $\Gamma\to\infty$. In this limit, the second term of eq.~(24) of the \emph{Letter} dominates and the minimization over $z_a$ and $z_b$ leads to the solutions $z_a=z_b=u/2$. Putting this back in the expression for the scgf and using the explicit expression for the scgf of a semi-infinite line SSEP (eq.~(3) of the \emph{Letter}), we obtain
\begin{equation}
\mu_{\text{inf}}(\lambda,\rho_a,\rho_b)=\int_{-\infty}^\infty\frac{\mathrm{d}k}{\pi}\log{\big[1+\omega(\lambda,\rho_a,\rho_b)\,\mathrm{e}^{-k^2}\big]}
\end{equation}
where we have used $\sinh^2u=\omega(\lambda,\rho_a,\rho_b)$. This is the scgf for the infinite line SSEP which has been well-studied in the literature.

\bibliography{letter}